**Phase-Controlled Epitaxy and Anisotropic Antiferromagnetism of Polar Wurtzite MnTe**

*Janusz Sadowski*[1,2] *, *Jaroslaw Z. Domagala*[1], *Piotr Dziawa*[1,3], *Anna Kaleta*[1], *Sania Dad*[1], *Maciej Wójcik*[2], *Dorota Janaszko*[1], *Sławomir Kret*[1], *Oleksii Liubchenko*[1], *Maciej Sawicki*[1], *Katarzyna Gas*[4,5,1] *

(1) Institute of Physics, Polish Academy of Sciences, Aleja Lotnikow 32/46, PL-02668 Warsaw, Poland

(2) Centre of Excellence ENSEMBLE3 Sp. z o.o., ul. Wolczynska 133, 01-919 Warsaw, Poland

(3) International Research Centre Magtop, Institute of Physics, Polish Academy of Sciences, Aleja Lotnikow 32/46, PL-02668 Warsaw, Poland

(4) Center for Science and Innovation in Spintronics, Tohoku University, 2-1-1 Katahira, Aoba-ku Sendai 980-8577, Japan

(5) Laboratory for Nanoelectronics and Spintronics, Research Institute of Electrical Communication, Tohoku University, 2-1-1 Katahira, Aoba-ku, Sendai 980-8577, Japan

janusz.sadowski@ifpan.edu.pl; gas.katarzyna.a2@tohoku.ac.jp

Altermagnetic spintronics requires materials in which compensated magnetic order, symmetry-controlled electronic responses, and epitaxial tunability can be combined in experimentally accessible thin films. MnTe is a key material in this context, but experimental studies have focused mainly on the stable NiAs-type polymorph, whereas the polar wurtzite phase remains largely unexplored. Here, we demonstrate molecular-beam epitaxy of nearly phase-pure wurtzite MnTe directly on GaAs(111)B and show that small changes in growth conditions strongly modify the phase composition, from a multiphase state with endotaxial NiAs-type inclusions to an almost single-phase polar wurtzite layer. High-resolution X-ray diffraction and transmission electron microscopy establish the epitaxial relationship, lattice parameters, microstructure, and strongly reduced density of secondary phases in the optimized films. SQUID magnetometry further establishes the optimized wurtzite layers as antiferromagnetic, with $T_N \cong 58$ K, a large negative Curie-Weiss temperature of $\theta_{CW} \cong -420$ K, and an anisotropic, predominantly field-linear excess magnetization appearing below $T_N$. These results identify epitaxial polar wz-MnTe as an antiferromagnetic semiconductor platform for future microscopic and transport studies of competing magnetic configurations and symmetry-controlled responses in wurtzite Mn chalcogenides.

## 1. Introduction

Antiferromagnetic (AFM) spintronics aims to exploit compensated magnetic order for robust and ultrafast information technologies, but its practical implementation requires reliable detection and electrical readout of the antiferromagnetic state. This remains challenging in conventional AFMs because the absence of a net magnetization strongly limits direct magnetic readout. Altermagnets (ALMs) offer one route around this limitation by combining the compensated order and negligible stray fields of antiferromagnets with spin-split electronic bands that enable ferromagnet-like transport responses without a large net magnetization.[1,2] Owing to this unusual combination, ALM materials have attracted considerable attention as candidates for prospective applications in spintronic devices.[3,4,5] This functional perspective has recently been extended to multiferroic ALMs, where ferroelectric polarization may enable electric-field control of spin-dependent electronic structures.[6,7,8]

MnTe is a key test material in this context because it combines magnetic semiconducting behavior with rich crystallographic polymorphism. Its stable NiAs-type alpha phase is now regarded as a representative altermagnetic compound,[2,9-12] whereas the magnetic and electronic properties of other MnTe polymorphs remain much less explored. The polar wurtzite $\beta$ phase is particularly attractive because it combines tetrahedral Mn coordination, broken inversion symmetry, and possible strain- or electric-field tunability. These features make wz-MnTe a promising platform for testing magnetic configurations and symmetry-controlled responses predicted for wurtzite Mn chalcogenides, but nearly phase-pure epitaxial films suitable for such studies have not been available so far.

Among the known MnTe polymorphs, the NiAs-type $\alpha$ phase is the stable structure under ambient conditions, whereas the zinc-blende $\gamma$ phase has been stabilized mainly in epitaxial layers grown on suitable substrates.[13-19] Rock-salt structure ($\delta$ phase) is metastable and can be stabilized e.g. by pressure; whereas wurtzite β-MnTe to our knowledge has been reported only in nonhomogeneous MnTe thin films as small admixture of secondary crystalline phase regions.[20-27] The interest in wz-MnTe is further enhanced by recent theoretical studies of polar wurtzite Mn chalcogenides. Wz-MnTe has been discussed as a possible noncentrosymmetric altermagnetic semiconductor capable of generating symmetry-selected pure spin photocurrents.[28] The possible altermagnetic phase of wurtzite MnTe and its relativistic spin–momentum locking have also been considered theoretically.[29] Related wz-MnSe has been experimentally realized and discussed in the context of altermagnetism,[30] and has recently been proposed theoretically as an altermagnetic multiferroic with sizable spin splitting and large spontaneous polarization.[31] More recent calculations comparing competing magnetic configurations have refined this picture by predicting a spin-degenerate stripe-type AFM ground state for pristine wurtzite MnX chalcogenides (X = S, Se, Te), together with nearby competing magnetic configurations that may be influenced by strain, magnetic history, temperature, or chemical modification.[32] At the same time, the polar wurtzite symmetry and the piezoelectric character typical of this structure make wz-MnTe attractive for strain- or electric-field tuning, and possibly for multiferroic functionality. Depending on the magnetic configuration, polar wurtzite Mn chalcogenides are also predicted to show distinct linear and nonlinear transport responses.[32] The nearly phase-pure wz-MnTe films reported here therefore fill an important experimental gap, providing a timely platform for benchmarking the magnetic response of this polar wurtzite polymorph and for testing the magnetic scenarios proposed for wurtzite Mn chalcogenides.

Here, we report molecular-beam epitaxy of nearly phase-pure $\beta$-MnTe thin films directly on GaAs(111)B substrates. By modifying the growth temperature under otherwise identical flux conditions, we switch the material between a multiphase $\beta/\alpha$-MnTe layer and an almost single-phase wurtzite $\beta$ film. High-resolution X-ray diffraction and transmission electron microscopy establish the phase purity, epitaxial relationship, lattice parameters, and nanoscale microstructure of these layers. Magnetometry further identifies the optimized wz-MnTe films as antiferromagnetic, with an anisotropic field-linear response emerging below Néel temperature ($T_N$). These results provide an experimental basis for future microscopic and transport studies of polar wurtzite Mn chalcogenides.

## 2. Results and discussion

### 2.1. MBE growth

Our samples have been grown in a home-built MBE system dedicated to Mn-doped narrow bandgap IV-VI semiconductors.[33] MnTe layers are grown directly on epi-ready GaAs(111)B substrates, glued to the molybdenum substrate holders by an eutectic Ga-In mixture.

Recently the MBE growth of pure $\alpha$-MnTe thin films deposited directly on GaAs(111)A substrates has been reported[34], but, to our knowledge the direct growth on GaAs(111)B has not been investigated yet. Interestingly Chen *et. al.*[35] investigated the MBE growth of MnTe on InP(111) substrates with A and B polarities. They reported nickeline $\alpha$ phase on InP(111)A and zinc-blende $\gamma$ phase on InP(111)B.[35]

In our MBE-growth processes standard resistively heated effusion cells have been used to generate $Te_2$ and Mn fluxes. The $Te_2$ to Mn flux ratio during the growth was chosen to about 5, the growth rate amounted to 0.02 ML/s. The two representative samples discussed here have been grown under the same Mn and Te fluxes and differ only in the growth temperature, which was changed by 50 °C (lower for mixed phase sample S#1, higher for almost pure wz phase sample S#2). This comparison shows that the MnTe polymorph selection on GaAs(111)B is highly sensitive to the growth temperature within otherwise unchanged MBE conditions. The MBE growth was monitored by reflection high energy electron diffraction (RHEED).

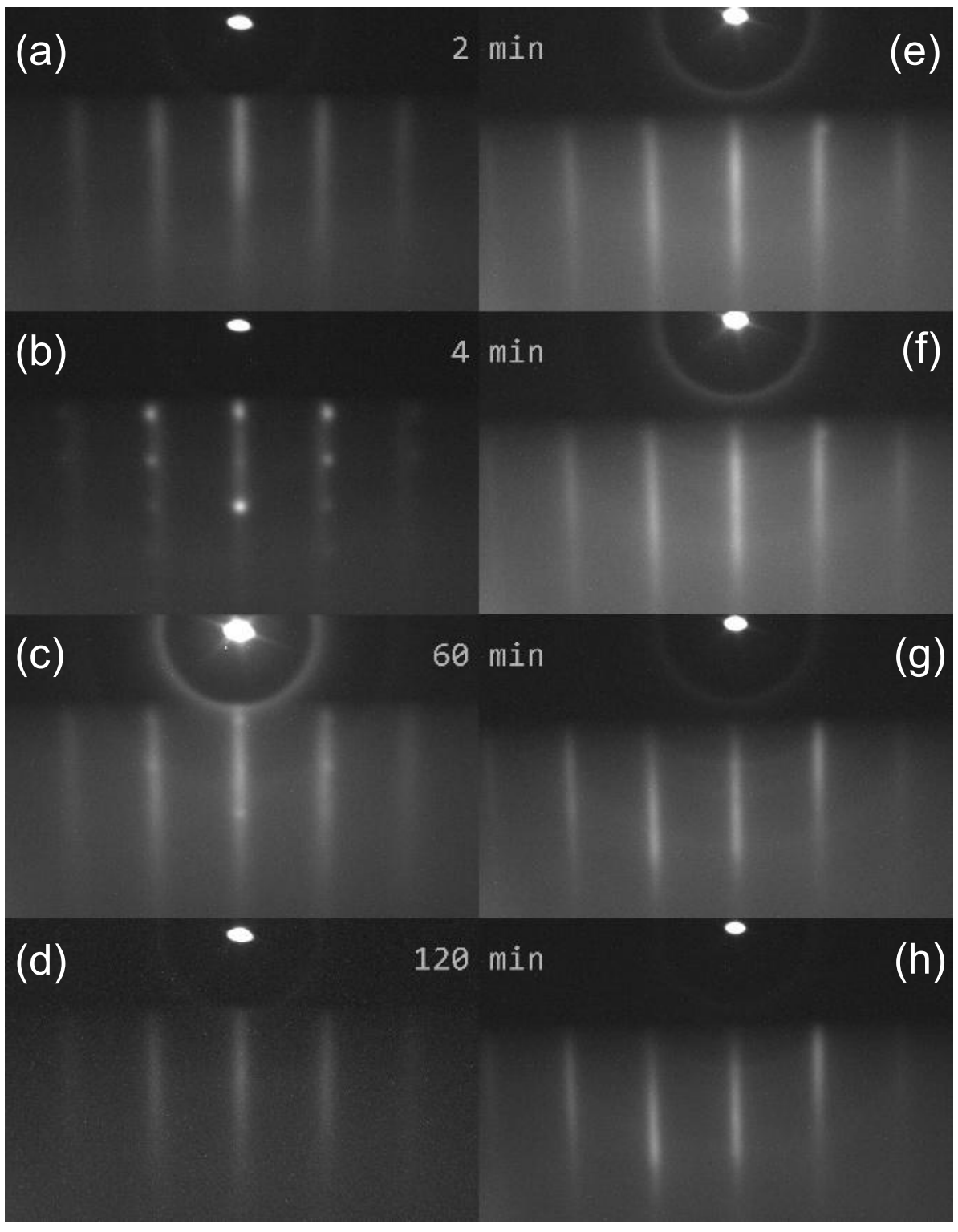

**Figure 1.** Evolution of RHEED images detected in different phases of the MBE growth. **(a) – (d)** multiphase wurtzite $\beta$-MnTe layer with nickeline $\alpha$-MnTe inclusions; **(e) - (h)** pure wurtzite MnTe. The occurrence of additional spotty patterns slightly out-ward dominating diffraction streaks visible in panels (b) and (c) indicates the presence of an $\alpha$-MnTe crystalline phase admixture.

The RHEED patterns recorded during growth reveal two distinct regimes, as shown in **Figure. 1**. In the multiphase sample, in the first stages of the MBE growth streaky diffraction from the dominant two-dimensional wurtzite layer coexists with additional spotty diffraction features, indicating the formation of three-dimensional crystalline inclusions. These spots remain aligned with the wurtzite streaks but are slightly shifted outward, consistent with a coherent secondary phase with a smaller in-plane lattice parameter. As confirmed below by X-ray diffraction and TEM, this secondary phase corresponds to a nickeline $\alpha$-MnTe, whereas the streaky pattern originates from the wurtzite $\beta$-MnTe matrix. In the optimized sample, the spotty features are strongly suppressed, indicating growth of an almost single-phase wurtzite layer.

### 2.2. X-ray diffraction results

High-resolution X-ray diffraction confirms the growth-dependent phase selection inferred from RHEED. Wide-range 2θ–ω scans shown in **Figure 2** reveal either a multiphase $\beta/\alpha$-MnTe layer or a nearly phase-pure wurtzite film, depending on the growth temperature. In the optimized sample (S#2), the additional $\alpha$-MnTe reflections are almost completely suppressed, apart from a weak residual 0004 peak, indicating a strongly reduced amount of the NiAs-type phase (see Figure 2b). The wurtzite MnTe layer is epitaxially aligned with GaAs(111)B, with the (0001) planes parallel to the substrate surface and no measurable tilt within the experimental uncertainty.

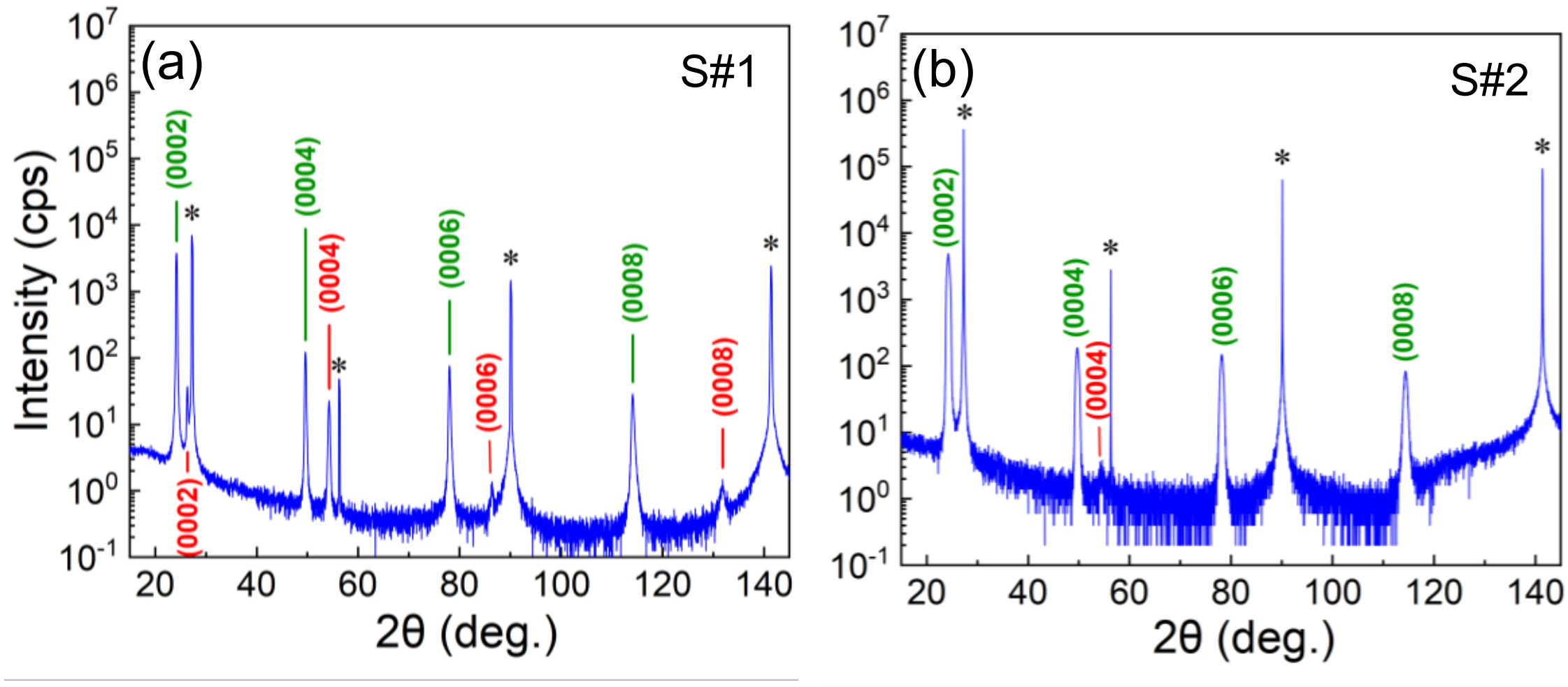

**Figure 2.** Wide 2θ/ω scans and rocking curves of 0002 Bragg reflections. **(a)** multiphase S#1 film, **(b)** near single phase MnTe film (S#2) grown by MBE on GaAs (111)B. Substrate reflections are marked with asterisks; MnTe wurtzite reflections are indicated in green, hexagonal (NiAs-type) reflections – in red.

The lattice parameters of the wurtzite $\beta$-MnTe phase are determined from symmetric and asymmetric Bragg reflections, yielding $a = 4.472 \pm 0.004$ Å and $c = 7.3420 \pm 0.0005$ Å.

The corresponding lattice parameters for bulk nickeline $\alpha$-MnTe are equal to:

$a = 4.1483$ Å $c = 6.7116$ Å.[36]

In our samples the lattice parameters of $\alpha$-MnTe inclusions calculated from the angular positions of 0004 Bragg reflections amount to:

$a = 4.152$ Å, $c = 6.748$ Å (S#1)

$a = 4.144$ Å, $c = 6.7597$ Å (S#2)

These values indicate that the residual $\alpha$-MnTe inclusions are locally distorted within the $\beta$-MnTe matrix, with the largest deviation observed along the out-of-plane direction.

The MnTe lattice parameters measured by us are consistent with those previously reported, but have been measured here with significantly improved precision, reflecting the high (as for MnTe films) structural quality of the layers. Rocking curve measurements show full widths at half maxima of $\beta$-MnTe (see **Figure S1** and **Figure S2** in Supplementary Information) on the order of (~0.3–0.5°), indicating a finite mosaic spread typical for heteroepitaxial systems and suggesting the presence of dislocation-related disorder. To further assess the defect structure, reciprocal space maps were analysed using models developed for III-nitride wurtzite materials. The results indicate misfit dislocation densities on the order $\sim 10^8$ cm$^{-2}$ (see **Figure S3** in Supplementary Information).

These estimates should be treated as semi-quantitative, as the applied models are adapted from III-nitride systems. A detailed analysis of peak shapes, line broadening, and dislocation modelling is provided in the Supplementary Information.

### 2.3. TEM investigations

Transmission electron microscopy provides direct insight into the nanoscale origin of the phase selectivity. In the multiphase sample S#1, the $\beta$-MnTe matrix is locally interrupted by $\alpha$-MnTe inclusions located at the GaAs(111)B/$\beta$-MnTe interface (see **Figure 3**). These inclusions are not randomly distributed throughout the layer, but are rooted at the interface and extend along the [0001] growth direction. High-resolution scanning transmission electron microscopy (STEM) images confirm the predominance of the wurtzite structure in the surrounding matrix, while local zinc-blende stacking segments indicate additional tetrahedral polytypism within the $\beta$-MnTe layer.

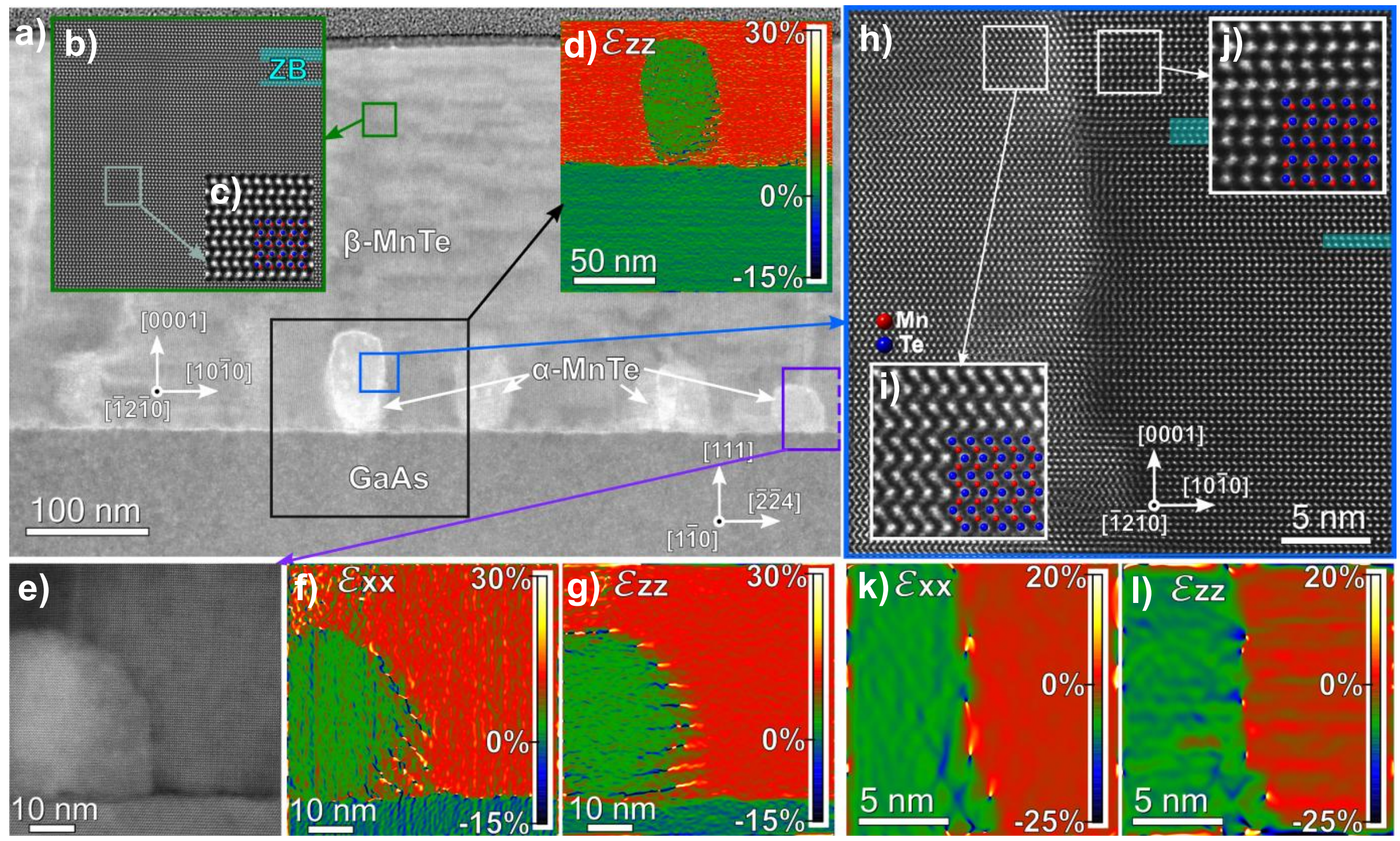

**Figure 3.** **(a)** Scanning transmission electron microscopy (STEM) image of a cross-section of sample S#1 containing substantial amount of $\alpha$-MnTe inclusions at GaAs(111)B/$\beta$-MnTe interface. **(b)** Magnified image of the region highlighted by the green square, showing the wurtzite phase of MnTe with local zinc blende segments marked in cyan. **(c)** Magnified area marked by the grey square in (b), with the atomic model superimposed, confirming the wurtzite crystal structure viewed along the [1$\bar{2}$10] zone axis. **(d)** Geometric phase analysis (GPA) strain map showing the $\varepsilon_{zz}$ component around an $\alpha$-MnTe inclusion marked by the black square in (a). **(e)** Magnified STEM image of the $\alpha$-MnTe inclusion marked by the purple square in (a). **(f, g)** GPA strain maps of the region shown in (e), presenting $\varepsilon_{xx}$ and $\varepsilon_{zz}$ components, respectively, with GaAs used as the reference. Misfit dislocations are visible at the interfaces between the phases. **(h)** High-resolution STEM image of the interface between $\alpha$-MnTe and $\beta$-MnTe phases, marked with blue square in Fig.3a. **(i, j)** - Magnified regions marked in (h) with superimposed atomic models, confirming $\alpha$-MnTe and $\beta$-MnTe crystal structures. **k,l)** GPA strain maps of the region shown in (h), presenting $\varepsilon_{xx}$ and $\varepsilon_{zz}$ components, respectively.

Geometric phase analysis (GPA) shows that the $\alpha$-MnTe inclusions are accompanied by localized strain fields and misfit dislocations at the phase boundaries. This indicates that the secondary phase acts as a local strain-relaxation pathway during the early stage of growth. The optimized sample S#2 with TEM cross-section displayed in **Figure 4** shows a markedly different microstructure: $\alpha$-MnTe is barely detectable by X-ray diffraction and appears in TEM only as sparse nanoscale inclusions at the GaAs(111)B/$\beta$-MnTe interface. Thus, phase purification of wz-MnTe proceeds mainly through suppression of interfacial $\alpha$-MnTe nucleation rather than through elimination of randomly distributed bulk precipitates.

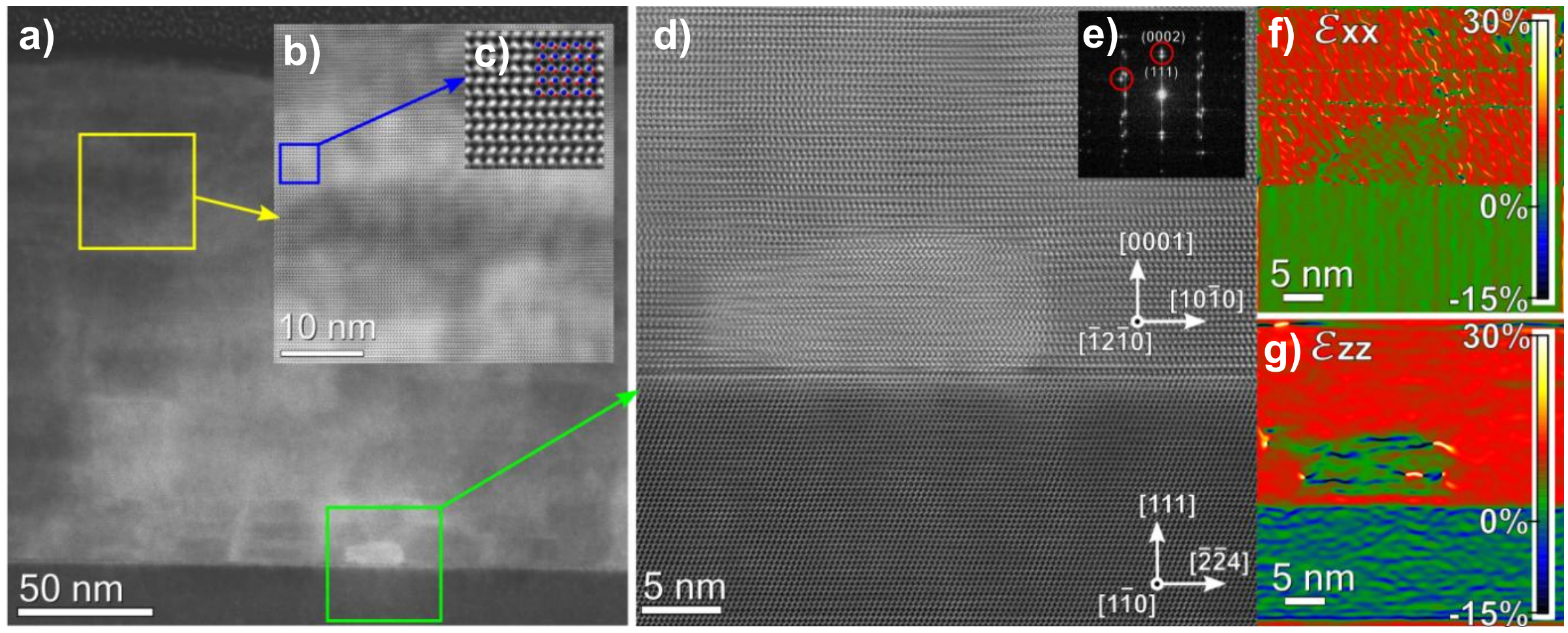

**Figure 4.** **(a)** Cross-sectional STEM image of sample S#2 containing an almost phase-pure $\beta$-MnTe layer grown on GaAs(111)B. Only a few small $\alpha$-MnTe inclusions are observed at the GaAs(111)B/$\beta$-MnTe interface. **(b)** Magnified image of the region marked by the yellow square in (a), showing the β-MnTe layer. **(c)** The inset shows an enlarged atomic-resolution image from area marked with blue square, with the corresponding atomic model superimposed. **(d)** High-resolution STEM image of the interface region marked by the green square in (a), showing a small $\alpha$-MnTe inclusion embedded at the interface between GaAs and $\beta$-MnTe. **(e)** The inset shows the corresponding FFT with two reflections chosen for GPA analysis. **(f, g)** GPA strain maps of the region shown in (c), presenting $\varepsilon_{xx}$ and $\varepsilon_{zz}$ components, respectively, highlighting local strain accumulation around the $\alpha$-MnTe inclusion and misfit dislocations at the phase boundaries.

The details concerning calculations of misfits between crystalline lattices of GaAs(111)B substrate, $\beta$-MnTe epilayer and $\alpha$-MnTe precipitates and presented in the Supplementary Information.

## 2.4. Magnetic measurements

Magnetization measurements are performed using an MPMS XL SQUID magnetometer in magnetic fields up to $\mu_0 H = 4$ T and in the temperature range from 2 to 360 K. The samples are measured with the magnetic field applied either parallel or perpendicular to the film plane, corresponding to $H \perp c$ and $H \| c$, respectively. Because the MnTe signal is small compared with the GaAs substrate contribution, the magnetic moment is corrected by subtracting the response of a reference GaAs(111)B substrate piece prepared, mounted, and measured using the same protocol. Details of the subtraction procedure and the control of orientation-dependent artifacts are given in Supplementary Information (including references[37-45]). The data discussed below are shown for the optimized sample S#2, for which the correction procedure allows a quantitative Curie-Weiss analysis.

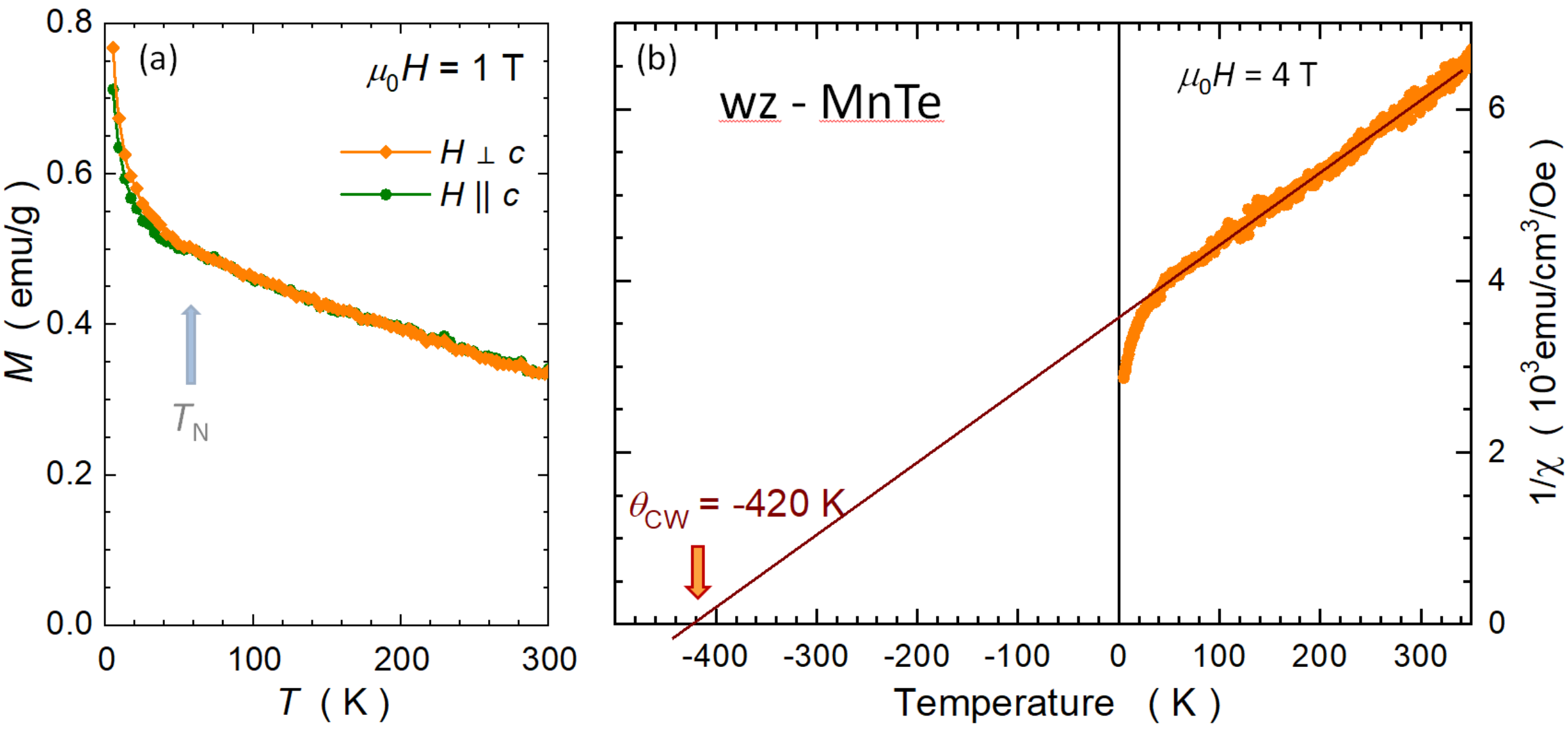

**Figure 5. (a)** Temperature $T$ dependence of the magnetization $M$ of MnTe/GaAs(111)B (sample S#2) established from measurements at $\mu_0 H = 1$ T for magnetic field applied parallel and perpendicular to the wurtzite $c$ axis. **(b)** Temperature dependence of the inverse of magnetic susceptibility $\chi^{-1} = H/M$ obtained from measurements at $\mu_0 H = 4$ T applied perpendicular to the wurtzite $c$ axis. The solid line is the linear interpolation to the Curie-Weiss law.

The magnetometric data provide the first indication of magnetic ordering in the wurtzite MnTe layer. **Figure 5a** shows $M(T)$ measured at $\mu_0 H = 1$ T for the magnetic field applied parallel and perpendicular to the wurtzite $c$ axis. In both configurations, the response is dominated by a Curie-Weiss-like paramagnetic contribution, which increases continuously upon cooling. Below about 60 K, $M(T)$ deviates from this high-temperature trend: an additional low-temperature increase develops, a weak anomaly appears, and the two curves measured in different field orientations start to bifurcate. We assign this feature to the transition from the paramagnetic state to the antiferromagnetically ordered state. A more accurate determination, based on the deviation from the Curie-Weiss behavior and on the onset of magnetic anisotropy, gives $T_N = 58 \pm 2$ K, as shown in **Figure 6**.

Figure 5b presents the inverse susceptibility, $1/\chi = H/M$, measured at $\mu_0 H = 4$ T. Between approximately 70 and 350 K, $1/\chi(T)$ follows a linear Curie-Weiss dependence without any visible deviation. The extrapolation gives a negative Curie-Weiss temperature $\theta_{CW} = -420 \pm 50$ K, while the Curie constant yields an effective spin $S = 2.6 \pm 0.2$. This value is consistent, within the experimental uncertainty, with the high-spin $Mn^{2+}$ configuration, $S = 5/2$. The large negative $\theta_{CW}$ demonstrates dominant antiferromagnetic exchange interactions in wz-MnTe. The resulting frustration index, $f = |\theta_{CW}|/T_N$, is close to 7. This value shows that the ordering temperature is substantially reduced with respect to the exchange scale, although it remains below the range usually associated with strongly frustrated magnets.[46] For comparison, $f$ is about 9 for zb-MnTe, using $\theta_{CW} \approx -530$ K and $T_N \approx 60$ K,[47,48,49] whereas for NiAs-type α-MnTe it is about 2, using $\theta_{CW} \approx -585$ K and $T_N \approx 307$ K.[50,36] Thus, the value obtained for wz-MnTe is much closer to that of the metastable zinc-blende polymorph than to the stable NiAs phase. This is physically reasonable, since wz-MnTe and zb-MnTe are both tetrahedrally bonded polymorphs with similar local superexchange pathways, whereas α-MnTe has a distinct octahedrally coordinated exchange network. The low $T_N$ of wz-MnTe therefore does not reflect weak antiferromagnetic exchange, but rather a reduced ordering efficiency characteristic of tetrahedrally coordinated MnTe polymorphs. This observation is also relevant in view of recent calculations for the wz-MnX family, which predict competing exchange interactions and place the collinear stripe-AFM configuration close in energy to a noncollinear all-in-all-out state[32].

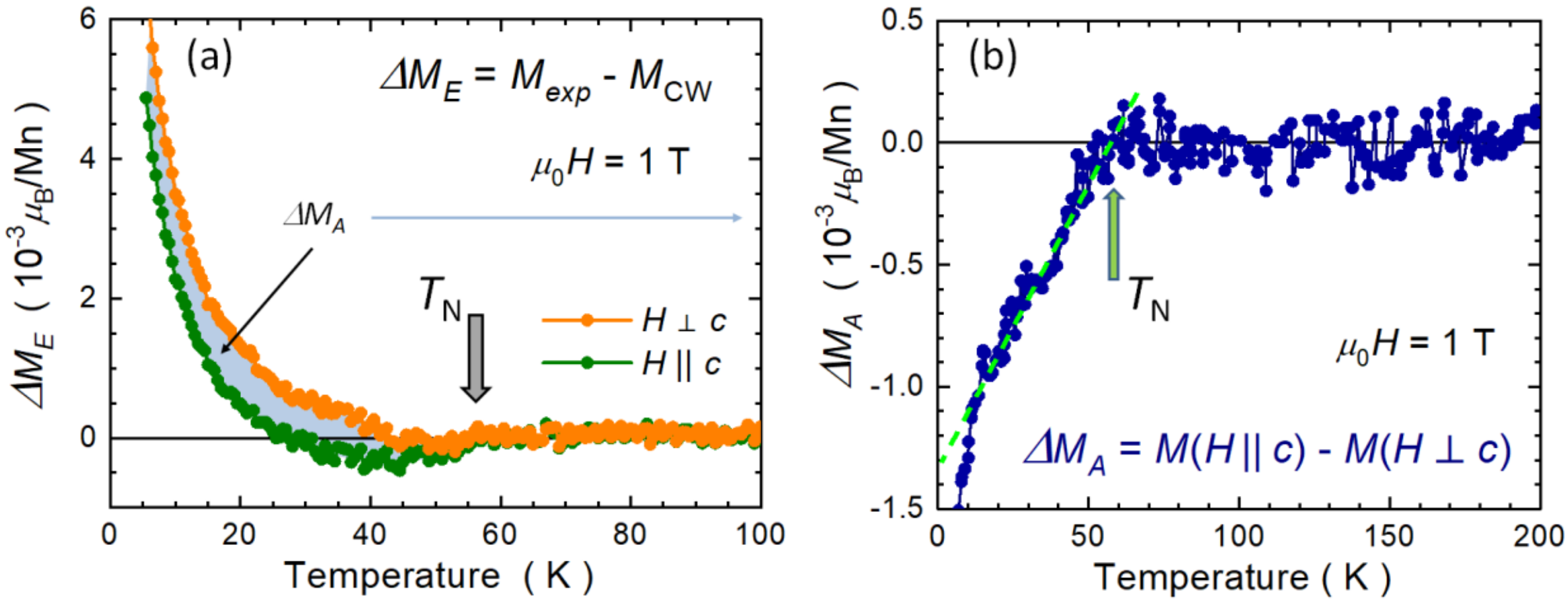

**Figure 6.** **(a)** Excess magnetization established below the Néel point, $\Delta M_E = M_{exp} - M_{CW}$, obtained after subtracting the Curie-Weiss contribution obtained from the linear approximation of inverse magnetic

susceptibility data in Fig. 5 (b). The pale blue marked area indicates the magnetic anisotropy which is shown in the right panel. **(b)** Magnetization anisotropy defined as $\Delta M_A = M(H||c) - M(H\perp c)$. Both $\Delta M_E$ and $\Delta M_A$ vanish at the same $T_N = 58 \pm 2$ K.

We now turn to the magnetic response below $T_N$. The effective spin close to the expected $Mn^{2+}$ value argues against assigning the low-temperature increase of $M(T)$ to a small concentration of free Mn spins only. Instead, it indicates that the anomaly is connected with the collective response of the Mn sublattice. To isolate the contribution emerging in the ordered phase from the dominant Curie-Weiss background, we subtract from the experimental magnetization, $M_{exp}$, the magnetization calculated from the high-temperature fit, $M_{CW}(T) = H \cdot C/(T - \theta_{CW})$. The parameters $C$ and $\theta_{CW}$ are obtained from the slope and intercept of the linear approximation to $1/\chi(T)$ in the high-temperature range, shown in Figure 5 (b). The resulting $\Delta M_E = M_{exp} - M_{CW}$ is shown in Figure 6 (a). This representation reveals a clear positive excess magnetization emerging below $T_N$ and reaching approximately $5 \times 10^{-3}$ $\mu_B$/Mn at 5 K. Importantly, $\Delta M_E$ vanishes at $58 \pm 2$ K, the same temperature at which the bifurcation of the two field orientations disappears. This shows that $\Delta M_E$ is not a simple continuation of the Curie-Weiss-like paramagnetic response, but an additional magnetic component associated with the onset of the antiferromagnetic state.

At the same time, the different magnitudes of the low-temperature response in the two field orientations show that the ordered magnetic phase is anisotropic. Below $T_N$, the magnetic response is larger for $H\perp c$ than for $H||c$. This anisotropy is highlighted in Figure 6 (a) by the pale blue background and shown explicitly in Figure 6 (b) as $\Delta M_A = M(H||c) - M(H\perp c)$. Within the experimental scatter, this difference decreases to zero at $T_N = 58 \pm 2$ K, consistent with the onset of $\Delta M_E$ at the magnetic transition.

The negative sign of $\Delta M_A$ shows that the susceptibility is larger for fields applied in the wurtzite *ab* plane ($H\perp c$) than for fields applied along the *c* axis. This establishes the ordered phase as magnetically anisotropic, but it does not by itself determine the orientation of the antiferromagnetic order parameter. In a simple collinear antiferromagnet below the spin-flop field, the susceptibility is generally larger for fields applied perpendicular to the Néel vector than for fields applied parallel to it. Therefore, a direct interpretation of the present anisotropy in terms of a single collinear Néel vector would suggest an order parameter closer to the *c* axis than to the in-plane field direction. However, such a simplified interpretation may not be sufficient for wz-MnTe, where recent calculations for the wz-MnX family predict closely competing magnetic configurations, including a stripe-type AFM ground state and nearby noncollinear states.[32] Thus, the measured anisotropy should be treated as an experimental constraint on the ordered magnetic state rather than as a complete determination of the spin structure.

The microscopic origin of the field-linear excess response cannot be determined from magnetometry alone. The data do not show clear hallmarks of a conventional remanent weak-ferromagnetic state[14,51] the thermoremanent magnetization is very weak and does not show a critical onset at $T_N$, no clear ZFC/FC splitting is resolved, and $M(H)$ remains nearly field-linear. The onset of the excess response at $T_N$, its anisotropy, and its absence in tetrahedrally coordinated cubic zb-MnTe[49] suggest that it is linked to the lower, uniaxial symmetry of the wurtzite phase rather than to a generic AFM response of MnTe. Possible microscopic origins include the proximity of competing collinear and noncollinear AFM configurations predicted for wz-MnX compounds, a symmetry-dependent altermagnetic electronic response requiring transport or spectroscopic verification, or a weak bound magnetic polaron (BMP) related contribution enabled by uniaxial anisotropy.[43,52] In all cases, the present data establish the ordered wz-MnTe phase as an anisotropic antiferromagnetic state and define an experimental benchmark for future microscopic, transport, and symmetry-resolved studies of epitaxial polar wz-MnTe.

## 3. Conclusion

We demonstrate phase-controlled molecular-beam epitaxy of polar wurtzite MnTe directly on GaAs(111)B. Under otherwise identical Mn and Te fluxes, a moderate change in the growth temperature strongly changes the phase composition, from a multiphase $\beta/\alpha$-MnTe layer with endotaxial interfacial $\alpha$-MnTe inclusions to an almost single-phase wurtzite film. X-ray diffraction and transmission electron microscopy establish the epitaxial relationship, lattice parameters, and nanoscale suppression of secondary phases in the optimized layers. Magnetometry identifies the optimized wz-MnTe films as antiferromagnetic semiconductors with $T_N = 58 \pm 2$ K, strong antiferromagnetic exchange, and an anisotropic field-linear response emerging below the ordering temperature. These results establish epitaxial polar wz-MnTe as a materials platform for future microscopic and transport studies of competing magnetic configurations and symmetry-controlled responses relevant to altermagnetic spintronics.

## 4. Experimental Section/Methods

*MBE growth*

The samples are grown in a home-made MBE system equipped with conventional resistively heated sources for Te and Mn (from MBE komponenten), rotatable substrate manipulator with the heater, 15 kV Staib RHEED gun and quartz microbalance for precise deposition rate calibrations. The partial pressure of residual gases was below $10^{-9}$ mbar in a standby MBE mode and at the level of $10^{-8}$ mbar during the growth. The MnTe growth was performed in Te-rich conditions, with $Te_2$/Mn flux ratio of about 5, at the substrate temperature in the range of 375 – 425 °C. At lower growth temperature more $\alpha$-MnTe rich samples were obtained. MnTe growth was performed on epi-ready GaAs(111)B substrates directly after thermal oxide desorption and lowering the substrate temperature from 590 °C to 375 – 425 °C (depending on the desired polytype of subsequently grown MnTe). MnTe growth rate, controlled by the Mn flux, amounted to about 0,25 Å/s

*X-ray diffraction measurements*

X-ray diffraction (XRD) measurements were performed using Empyrean (Malvern Panalytical) and Philips X'Pert MRD high-resolution diffractometers operating with $CuK\alpha_1$ radiation. High angular and spectral resolution was achieved using Ge(220)-based monochromators and mirror optics. Depending on the configuration, either a Pixcel 2D detector or a proportional detector was used.

Crystallographic reflections are denoted without parentheses throughout (e.g., 0002, 10-15).

*Transmission electron microscopy*

The structure and composition of the samples are investigated using an FEI Titan Cubed 80 300 transmission electron microscope (TEM) operating at 300 kV with an image aberration correction system. Cross-sections of the investigated samples are prepared using a Focused Ion Beam (FIB) lift–off procedure in a Helios Nanolab Dual beam 600 system with an Omniprobe nanomanipulator. The scanning transmission electron microscopy high-angle annular dark field detector (STEM−HAADF) images were acquired at the scattering angle ranges: 25 – 150 mrad or 80 – 200 mrad with a converged semi-angle of 9.5 mrad of the incident beam. Geometric phase analysis of acquired images was performed using the Gatan Digital Micrograph software. All TEM measurements are performed at room temperature.

*SQUID magnetometry*

Magnetization measurements are performed using an MPMS XL SQUID magnetometer in magnetic fields up to $\mu_0 H = 4$ T and in the temperature range from 2 to 360 K. The samples are measured in two orientations, with the magnetic field applied either parallel or perpendicular to the film plane.

The magnetic signal of the thin AF MnTe layer is small compared with the response of the GaAs substrate. Therefore, reliable extraction of the layer magnetization requires careful background subtraction, reproducible mounting, and closely matched measurement geometries. Small orientation-dependent artifacts or improper subtraction of the reference signal may otherwise lead to erroneous results comparable to the magnetic signal of interest.[37,38] The magnetic moment of each MnTe/GaAs specimen is therefore corrected by subtracting the signal of a reference GaAs(111) substrate piece from the same batch, prepared, mounted, and measured using the same protocol. All the relevant details of the experimental procedure are given in the paragraph S3 of Supplementary Information.

**Acknowledgements**

P. Dziawa acknowledge support by the "MagTop" project (FENG.02.01-IP.05-0028/23), carried out within the "International Research Agendas" program of the Foundation for Polish Science, co-financed by the European Union under the European Funds for Smart Economy 2021-2027 (FENG).

M. Sawicki and K. Gas gratefully acknowledge Piotr Nowicki for technical assistance with SQUID magnetometry measurements.

M. Wójcik and J. Sadowski are grateful for the „Centre of ExcelleNce for nanophotonicS, advancEd Materials and novel crystal growth-Based technoLogiEs - ENSEMBLE3" project which is carried out within the the 2.1 International Research Agendas programme of the Foundation for Polish Science co-financed by the European Union under the European Funds for Smart Economy 2021-2027 (FENG.02.01-IP.05-0044/24), project (MAB/2020/14), which is carried out within the International Research Agendas Programme (IRAP) of the Foundation for Polish Science co-financed by the European Union under the European Regional Development Fund and the Teaming Horizon 2020 programme (GA. No. 857543) of the European Commission and the project of the Minister of Science and Higher Education "Support for the activities of Centers of Excellence established in Poland under the Horizon 2020 program" under contract No. MEiN/2023/DIR/3797.

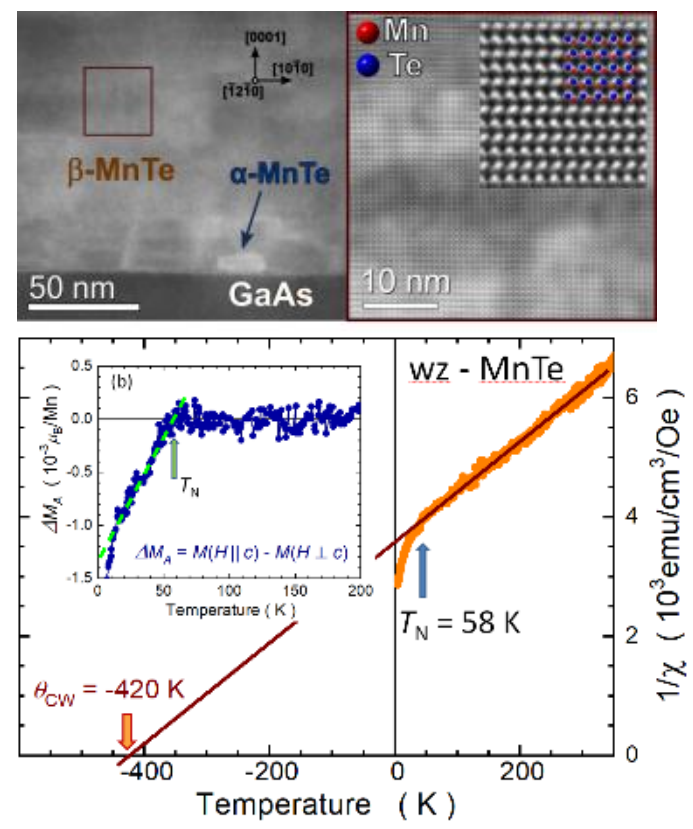

**TOC Figure**

**Phase-Controlled Epitaxy and Anisotropic Antiferromagnetism of Polar Wurtzite MnTe**

*Janusz Sadowski *, Jaroslaw Z. Domagala, Piotr Dziawa, Anna Kaleta, Sania Dad, Maciej Wójcik, Dorota Janaszko, Sławomir Kret, Oleksii Liubchenko, Maciej Sawicki, Katarzyna Gas**

# Supplementary Information

## S1. X-ray diffraction

Section S1 provides additional X-ray diffraction analysis supporting the structural characterization of the epitaxial wurtzite MnTe layers, including peak shape analysis, microstructural parameters, and defect estimates.

### S1.1 Rocking curves and peak shape analysis

The epitaxial alignment of the MnTe layers with the GaAs(111) substrate was verified by analyzing rocking curves measured for different azimuthal orientations (**Figure S1**). The MnTe(0001) planes {see panel (b)} are parallel to the GaAs(111) planes {see panel (a)}, with no measurable tilt within experimental uncertainty.

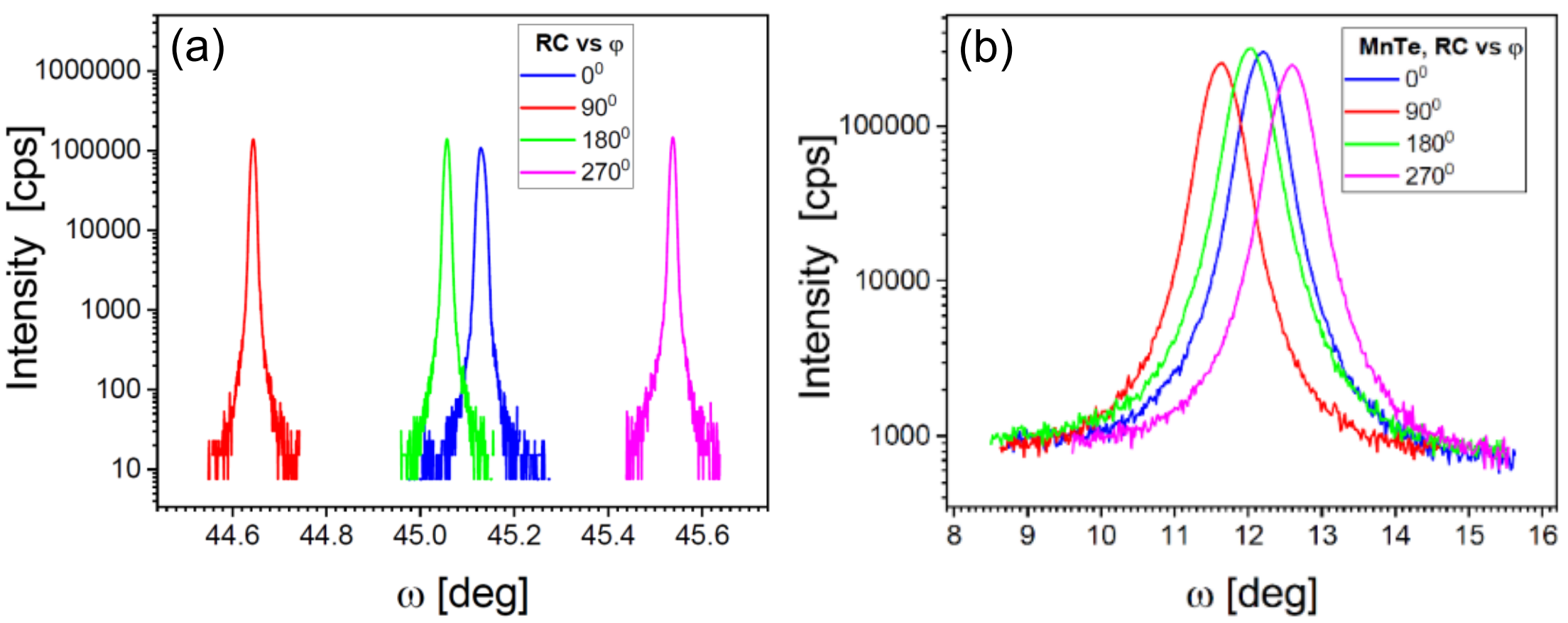

**Figure S1.** *Rocking curves of 333 GaAs reflection and 0002 of MnTe measured for selected azimuthal directions.*

Representative rocking curves for the MnTe-wz 0002 reflection are shown in **Figure S2** for both multiphase {panel (a)} and single-phase {panel (b)} samples. The single-phase wurtzite layer exhibits significantly narrower peaks, indicating improved crystalline quality.

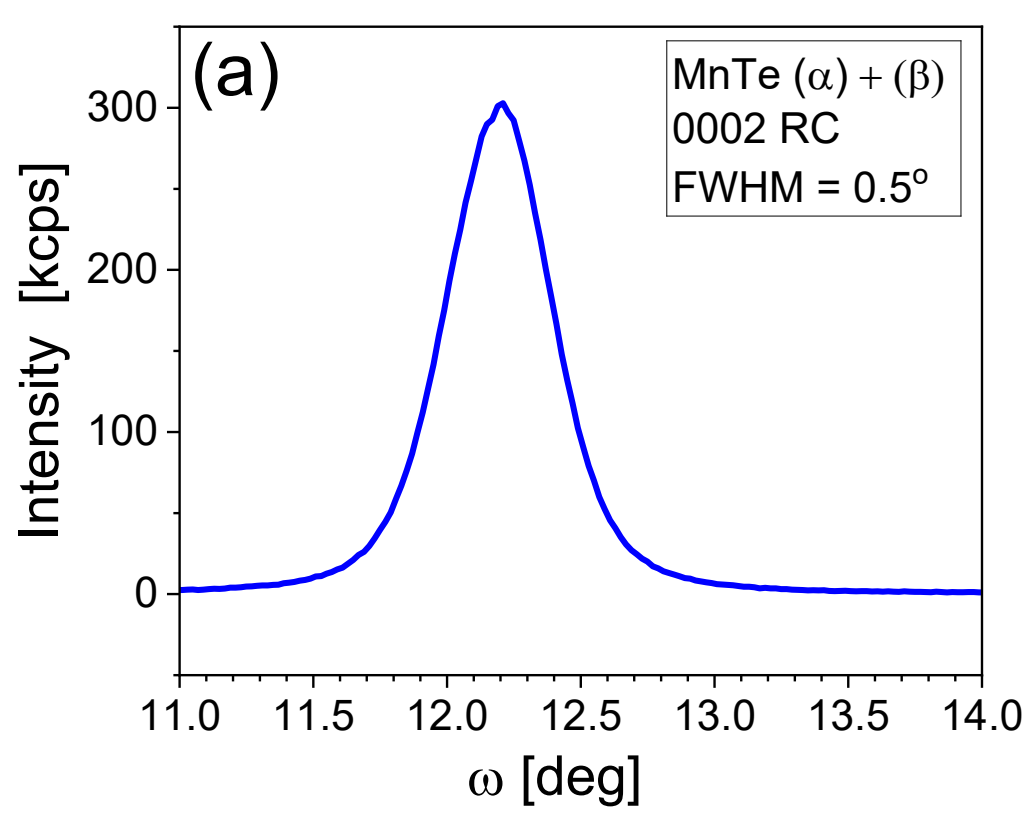

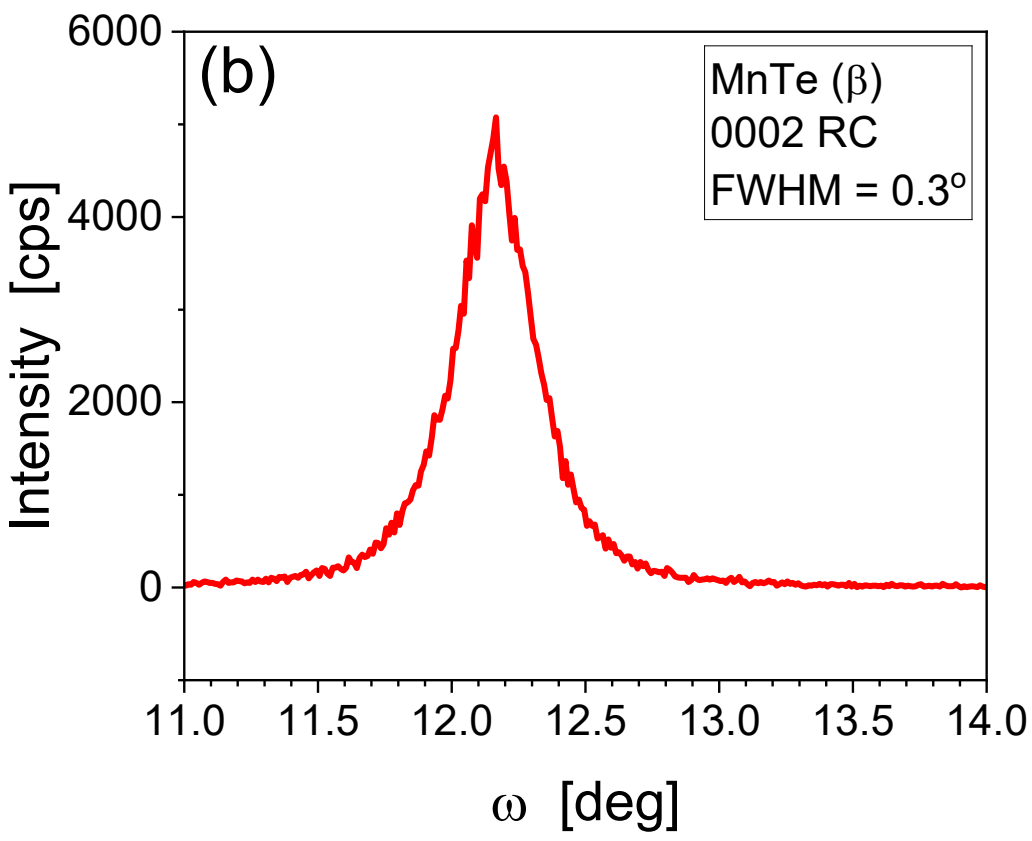

**Figure S2.** *Rocking curves of MnTe, 0002 for (a) – multiphase, and (b) - single-phase sample.*

The obtained lattice parameters are: **a = 4.472 ± 0.004 Å and c = 7.3420 ± 0.0005 Å,** and are consistent with the values reported in the main text.
The diffraction peaks were fitted using a pseudo-Voigt function to account for mixed Gaussian and Lorentzian broadening. The extracted full width at half maximum (FWHM) values were used as a robust measure of peak width in further analysis.
Tables S.T.1 and S.T.2 summarize the fitting parameters for the investigated samples.

## S1.2. Williamson–Hall analysis and microstructural parameters

The peak broadening was analyzed using the modified Williamson–Hall approach following Ratnikov et al [1]. The analysis was performed using the n = 2 representation, which provides a consistent description of the experimental data.
The extracted parameters indicate microstrain on the order of $\mathbf{10^{-3}}$–$\mathbf{10^{-2}}$ and characteristic block sizes in the sub-micrometer to micrometer range. No well-defined columnar or block-like structure is resolved within the sensitivity of the method.
Due to the mixed Gaussian–Lorentzian character of the diffraction peaks, these values should be treated as approximate and representative of the overall microstructural disorder.

## S1.3. Reciprocal space maps and dislocation modeling

Reciprocal space maps (RSMs) were measured around symmetric reflections to evaluate the defect structure of the MnTe-wz layers. Representative experimental and simulated maps are shown in **Figure S3**.

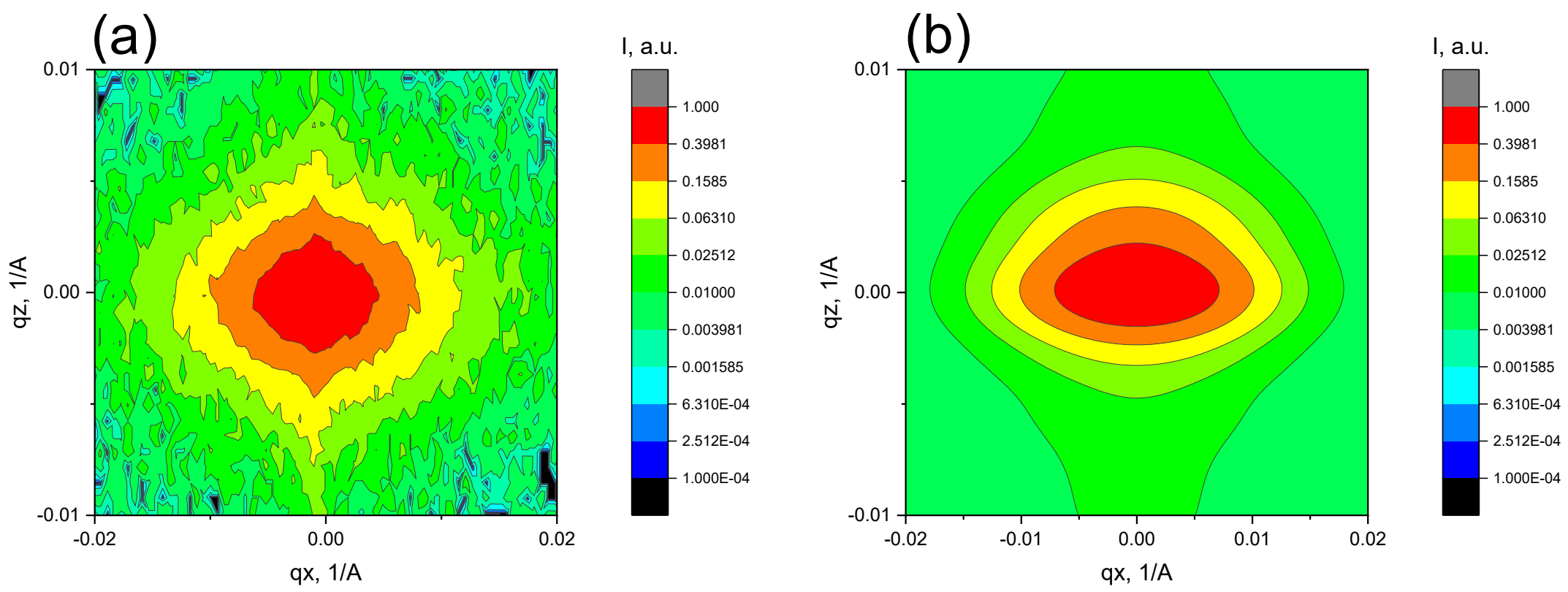

**Figure S3.** *(a) Experimental RSM of the MnTe ,0002 reflection. (b) Simulated RSMs including misfit dislocations.*

Simulations based on frameworks developed for III-nitride wurtzite materials were utilized to estimate the defect densities [2], indicating misfit dislocation densities on the order of ~$10^8$ cm$^{-2}$. While the agreement between experiment and simulation is qualitative—meaning these values should be regarded as order-of-magnitude estimates—the analysis does not explicitly account for stacking faults or α-MnTe inclusions, which may also contribute to the observed peak broadening. However, TEM data clearly reveal that defects, such as misfit dislocations, are concentrated at these α-MnTe inclusions. Consequently, the dislocation densities on the order of $10^8$ cm$^{-2}$ extracted experimentally from reciprocal space maps (RSMs) provide a macroscopically averaged measure of defects. This value represents an integral average over the dynamic X-ray beam footprint, which is projected from an initial incident beam size of 10 × 1 mm² ($v \times h$).

**Table S.T1.** Fitting parameters for sample 053-25 (diffractometer Empyrean, 2-bonce hybrid monochromator)

| **HKL #S1** | scan type | max position (deg) | intensity max (cps) | η | **FWHM** (deg) | area | **beta** integral (deg) | r_squared |
|---|---|---|---|---|---|---|---|---|
| "002" | RC | 12.146 | 611462 | 0.41 | 0.470 | 346089 | 0.566 | 0.9999 |
| "004" | RC | 24.852 | 23281 | 0.38 | 0.476 | 13503 | 0.580 | 0.9994 |
| "006" | RC | 39.057 | 18864 | 0.34 | 0.494 | 11354 | 0.602 | 0.9994 |
| "008" | RC | 57.125 | 10547 | 0.47 | 0.536 | 6993 | 0.663 | 0.9978 |
| "002" | 2θ/ω | 24.236 | 21703 | 0.69 | 0.060 | 1782 | 0.082 | 0.9947 |
| "004" | 2θ/ω | 49.637 | 481 | 0.66 | 0.113 | 72 | 0.150 | 0.9889 |
| "006" | 2θ/ω | 78.034 | 254 | 0.61 | 0.185 | 64 | 0.252 | 0.9689 |
| "008" | 2θ/ω | 114.156 | 87 | 0.87 | 0.321 | 40 | 0.455 | 0.9204 |

**Table S.T2.** Fitting parameters for sample 030-26 (diffractometer X-Pert MRD, 4-bonce monochromator + X-ray mirror)

| **HKL #S2** | scan type | max position (deg) | intensity max (cps) | η | **FWHM** (deg) | area | **beta** integral (deg) | r_squared |
|---|---|---|---|---|---|---|---|---|
| "002" | RC | 12.163 | 4649.5 | 0.68 | 0.341 | 2115.97 | 0.455 | 0.9971 |
| "004" | RC | 24.885 | 177.2 | 0.70 | 0.355 | 84.26 | 0.476 | 0.9969 |
| "006" | RC | 39.117 | 142.2 | 0.63 | 0.375 | 69.67 | 0.490 | 0.9966 |
| "008" | RC | 57.241 | 78.6 | 0.52 | 0.420 | 41.57 | 0.529 | 0.9921 |
| "002" | 2θ/ω | 24.275 | 111.1 | 0.65 | 0.069 | 10.39 | 0.094 | 0.9959 |
| "004" | 2θ/ω | 49.719 | 2.4 | 0.81 | 0.111 | 0.37 | 0.154 | 0.9470 |
| "006" | 2θ/ω | 78.183 | 1.4 | 0.94 | 0.182 | 0.35 | 0.247 | 0.8541 |
| "008" | 2θ/ω | 114.438 | 0.4 | 0.22 | 0.290 | 0.13 | 0.317 | 0.6045 |

# S2. TEM investigations

In the case of precipitates embedded in the matrix, the lattice distortion components $\varepsilon_{ij}^{GPA}$ can be related to local variation of the interplanar distances with respect to a chosen reference region by following equations

$$\varepsilon_{ij}^{GPA} = \frac{d_{loc} - d_{ref}}{d_{ref}}$$

where $d_{loc}$ denotes local interplanar distance, $d_{ref}$ corresponds to interplanar distance in the undeformed material region. This approaches can be applied to α-MnTe precipitate in β-MnTe matrix, grown on GaAs(111). In this system, three phases coexist. Due to the similarity of the space groups of the two hexagonal structures (α-MnTe and β-MnTe), $\{10\bar{1}0\}, \{\bar{2}110\}$, $\{0001\}$ planes are common to both lattices. However, for the cubic GaAs substrate, the corresponding planes must be selected carefully, see Table S2.T1. In this convention, in-plane x-direction corresponds to $[10\bar{1}0]$ for α and β-MnTe and $[22\bar{4}]$ for GaAs, while out-of-plane z-direction corresponds to $[0001]$ for α and β-MnTe and $[111]$ for GaAs.

Therefore, the misfits M between bulk α-MnAs, bulk β-MnTe (as measured by XRD), and GaAs can be defined with respect to GaAs and GaAs is either referenced to GaAs (**Figure 3 (d,f,g)** and **Figure 4 (f,g**) of the main manuscript or with respect to β-MnTe **Figure 3 ( k,l)**, ibid., as:

$$M = \frac{d_{MnTe}^{bulk} - d_{GaAs}^{bulk}}{d_{GaAs}^{bulk}}$$

However, residual strain components $\varepsilon_{ij}$ remain when full relaxation is not achieved:

$$\varepsilon_{ij} = \frac{d_{MnTe}^{loc} - d_{MnTe}^{bulk}}{d_{MnTe}^{bulk}}$$

where the local lattice parameter d (measured along a given crystallographic direction) in the strained MnTe region is referenced to the corresponding bulk (unstrained) MnTe value. Finally, residual strain in precipitate (or layer can be expressed in terms of the GPA-measured strain components $\varepsilon_{ij}^{GPA}$ obtained from HRTEM/HR-STEM images:

$$\varepsilon_{ij} = \frac{\varepsilon_{ij}^{GPA} - M}{1 + M}$$

**Table ST3**

| Corresponding crystallographic planes/directions | | GaAs | β-MnTe | α-MnTe |
|---|---|---|---|---|
| z-direction (out of plane) | distance d [Å] $(111)_{GaAs}/(0002)_{MnTe}$ | 3.264 | 3.671 | 3.356 |
| | Misfit to GaAs | - | 12.47 % | 2.81% |
| | Misfit to β-MnTe | -11.09% | - | - 8.59% |
| x-direction (in plane) | distance d [Å] $(22\bar{4})_{GaAs}/(30\bar{3}0)_{MnTe}$ | 1.154 | 1.291 | 1.198 |
| | Misfit to GaAs | - | 11.87 % | 3.81 % |
| | Misfit to β-MnTe | - 10.61% | - | - 7.20% |

# S3. Magnetometry protocol and background subtraction

Magnetization measurements are performed using an MPMS XL SQUID magnetometer in magnetic fields up to $\mu_0 H = 4$ T and in the temperature range from 2 to 360 K. The samples are measured in two orientations, with the magnetic field applied either parallel or perpendicular to the film plane.

The magnetic signal of the thin MnTe layer is small compared with the response of the GaAs substrate, as illustrated in **Figure S4**. Therefore, reliable extraction of the layer magnetization requires careful background subtraction, reproducible mounting, and closely matched measurement geometries. Small orientation-dependent artifacts or improper subtraction of the reference signal may otherwise lead to erroneous results comparable to the magnetic signal of interest [3,4]. The magnetic moment of each MnTe/GaAs specimen is therefore corrected by subtracting the signal of a reference GaAs(111) substrate piece from the same batch, prepared, mounted, and measured using the same protocol. The reference response is scaled to the mass of the MnTe/GaAs specimen before subtraction.

This procedure is necessary because the GaAs background is not strictly temperature independent. The magnetic purity of commercial substrates may vary between vendors and batches, and magnetically active impurities are known to affect not only oxide substrates [5,6], including MgO [7], but also GaAs [8]. In the present case, the GaAs reference signal changes by about 2% in the measured temperature range. This variation is small on the scale of the total substrate-dominated moment, but large on the scale of the extracted MnTe-layer contribution. Treating magnetic susceptibility of GaAs as temperature independent would distort $M(T)$ and would prevent recovery of the linear $1/\chi(T)$ Curie-Weiss dependence used in the main text.

Before magnetometry, the backside of each specimen is cleaned to remove the Ga-In glue used to attach the substrate to the molybdenum holder during MBE growth. Such metallic residues may produce nonlinear and sometimes hysteretic magnetic backgrounds [9]. Because the thin GaAs(111) substrates are mechanically fragile, no polishing or thinning was applied. The remaining metallic material is gently scraped using a white ceramic spatula to avoid magnetic contamination from metallic tools. Although this procedure removed most of the visible residue, a small Ga-In-related background cannot be completely excluded. For this reason, the quantitative Curie-Weiss analysis in the main text is performed using the $\mu_0 H = 4$ T data, where the nonlinear contribution from this residual metallic background is expected to be saturated.

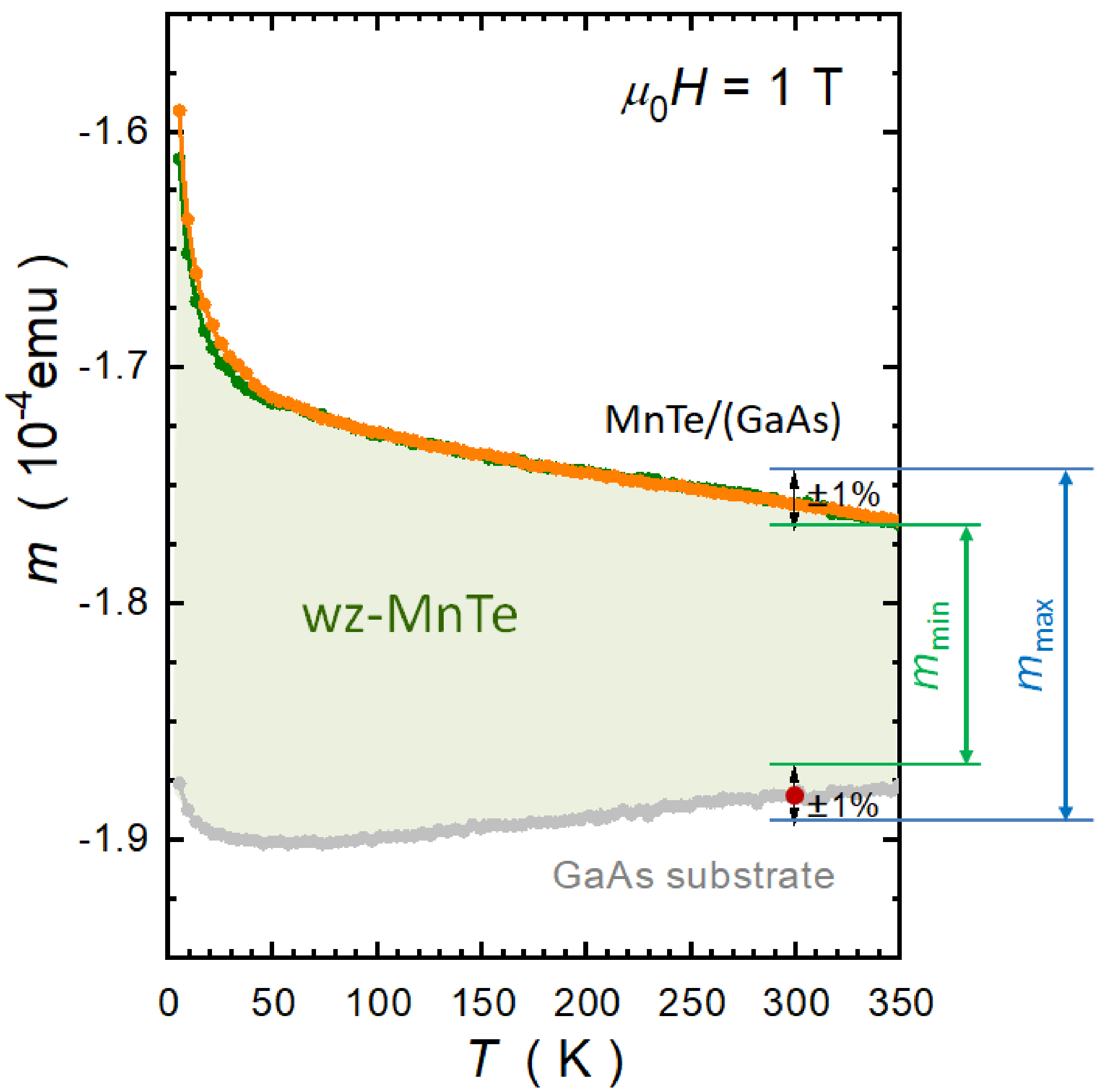

**Figure S4.** *Temperature dependence of the magnetic moment measured at $\mu_0 H$ = 1 T for the MnTe/GaAs(111)B Sample 2 and for a reference GaAs(111) substrate piece. The reference signal is scaled to the mass of the MnTe/GaAs specimen. The pale-green shaded region marks the MnTe-layer contribution obtained after subtraction. The arrows illustrate the range of extracted MnTe-layer responses that would result from systematic shifts of the large sample and reference signals by about 1% in opposite directions before subtraction. The geometrical coupling factor $\gamma$ = 1.02 [3] is included in the analysis.*

Another source of systematic error is the geometrical coupling between the specimen and the SQUID pick-up coils, which depends on sample shape and orientation [3,10]. For extended thin-plate specimens this coupling can differ by several percent between the two field orientations. In the present case, where the MnTe-layer signal is obtained by subtracting two large substrate-dominated moments, such an error would preclude a reliable quantitative comparison of the two geometries and could even change the apparent sign of the extracted layer contribution.

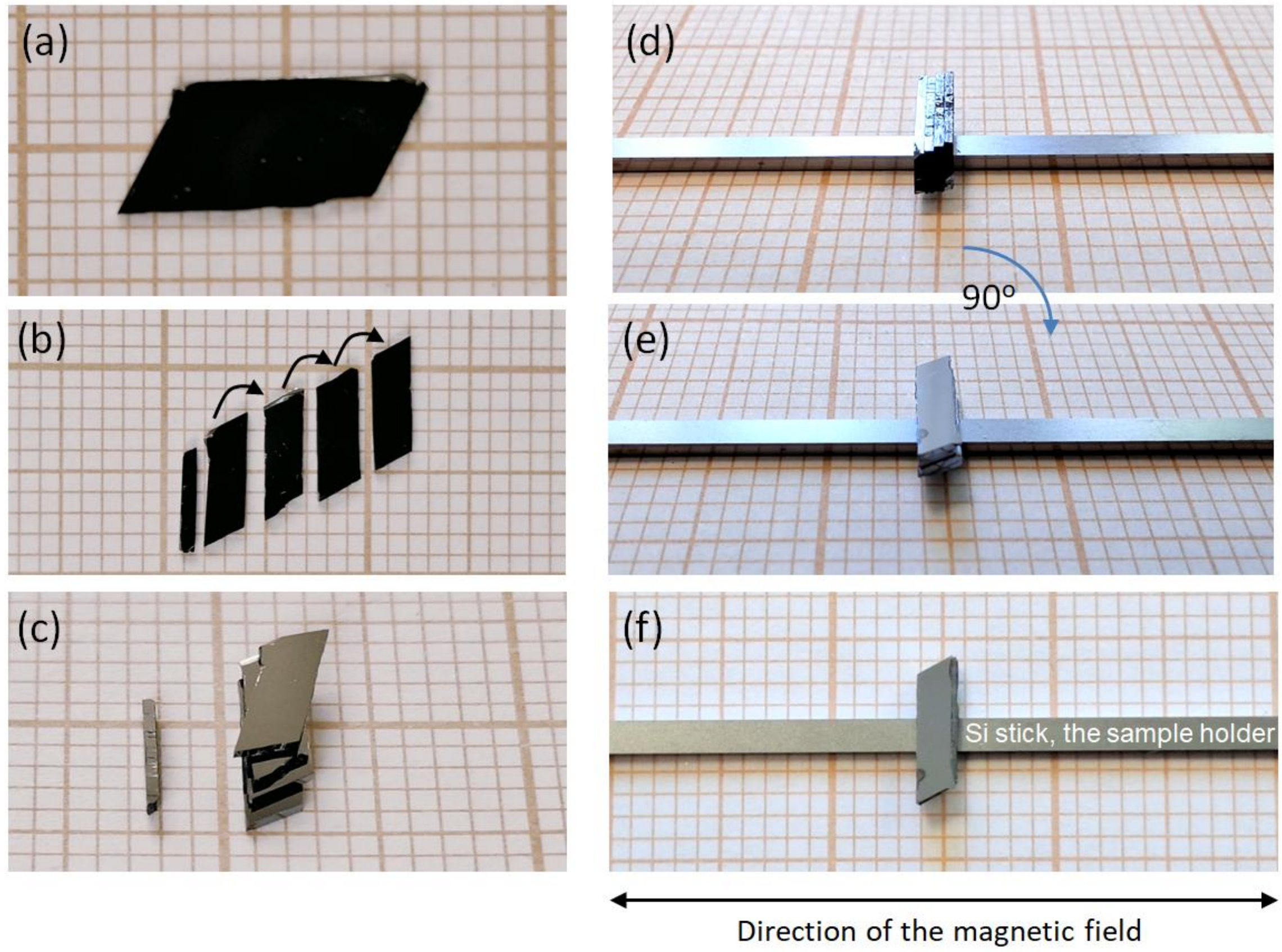

**Figure S5** *Specimen preparation method to minimize orientation-dependent effects in magnetometry. The original thin MnTe/GaAs piece (**a**) is cut into narrow strips (**b**),which are stacked (**c**), and after alignment on one edge they are glued into a rod-like cuboid with an approximately square cross-section. The two measurement geometries are obtained by rotating the same specimen from (**d** – H perpendicular to plane) by 90 degrees around its long axis (**e** – H in plane), keeping the effective shape seen by the SQUID pick-up coils nearly unchanged. (**f**) is a birds eye view of (e).*

To minimize this effect, the MnTe/GaAs pieces are reshaped into a rod-like geometry, as shown in **Figure S5**. The original thin sample piece is cut along easy cleavage directions into narrow strips about 1.8 mm wide. The strips are stacked and glued with diluted GE varnish [11], forming a cuboid with an approximately square cross-section of about $1.8 \times 1.7\ \mathrm{mm}^2$ and a length of about 5 mm. The two field orientations are obtained by rotating the same cuboid by 90 degrees around its long axis {Figure S5 (d-e)}, using the same holder. This keeps the effective shape seen by the SQUID pick-up coils nearly unchanged and reduces orientation-dependent changes of the coupling factor [12]. For the present specimens, $\gamma \cong 1.02$ [3] is included in the analysis.

After subtraction of the GaAs reference signal, the corrected moment is normalized to the MnTe layer volume. The sample analyzed quantitatively in the main text is selected because its residual background is sufficiently small and reproducible to allow a meaningful Curie-Weiss analysis. Other samples show the same qualitative behavior, including the low-temperature anomaly and orientation-dependent response, but larger residual backgrounds prevent equally reliable extraction of absolute Curie-Weiss parameters.

Figure S4 provides an internal consistency check of the procedure. Above about 60 K, the two field orientations give the same temperature dependence within the noise level, whereas the low-temperature anomaly and orientational difference appear only below the ordering temperature discussed in the main text. A systematic mismatch of only about 1% between the sample and reference moments, acting in opposite directions before subtraction, would change the extracted MnTe-layer magnetization by about ±20% at 300 K. This sensitivity justifies the matched reference measurement, special specimen geometry, and high-field analysis used here.